\documentclass[runningheads]{llncs}
\usepackage{amsfonts}
\usepackage{amssymb}
\usepackage{multirow}
\usepackage{graphicx}
\usepackage{cite} 
\usepackage{color}

\begin{document}
\title{Rethinking the optimization process for self-supervised model-driven MRI reconstruction} 

\author{Weijian Huang\inst{1,2,3}\and Cheng Li\inst{2}\and Wenxin Fan\inst{1,2} \and Yongjin Zhou\inst{4}\and Qiegen Liu\inst{5}\and Hairong Zheng\inst{2,3} \and Shanshan Wang\inst{2,3}}
\authorrunning{W. Huang et al.}
\institute{University of Chinese Academy of Sciences, Beijing, China \and Paul C. Lauterbur Research Center for Biomedical Imaging, Shenzhen Institutes of Advanced Technology, Chinese Academy of Sciences, Shenzhen, Guangdong, China
\and Pengcheng Laboratory, Shenzhen, Guangdong, China
\and Shenzhen University, Shenzhen, Guangdong, China
\and Nanchang  University, Nanchang, Jiangxi, China
\\
\email{Sophiasswang@hotmail.com, ss.wang@siat.ac.cn}}

\maketitle

\begin{abstract}
Recovering high-quality images from undersampled measurements is critical for accelerated MRI reconstruction. Recently, various supervised deep learning-based MRI reconstruction methods have been developed. Despite the achieved promising performances, these methods require fully sampled reference data, the acquisition of which is resource-intensive and time-consuming. Self-supervised learning has emerged as a promising solution to alleviate the reliance on fully sampled datasets. However, existing self-supervised  methods suffer from reconstruction errors due to the insufficient constraint enforced on the non-sampled data points and the error accumulation happened alongside the iterative image reconstruction process for model-driven deep learning reconstrutions. To address these challenges, we propose K2Calibrate, a K-space adaptation strategy for self-supervised model-driven MR reconstruction optimization. By iteratively calibrating the learned measurements, K2Calibrate can reduce the network’s reconstruction deterioration caused by statistically dependent noise. Extensive experiments have been conducted on the open-source dataset FastMRI, and K2Calibrate achieves better results than five state-of-the-art methods. The proposed K2Calibrate is plug-and-play and can be easily integrated with different model-driven deep learning reconstruction methods.
\end{abstract}

\section{Introduction}
\begin{figure}
\centering
\includegraphics[width=0.8\textwidth]{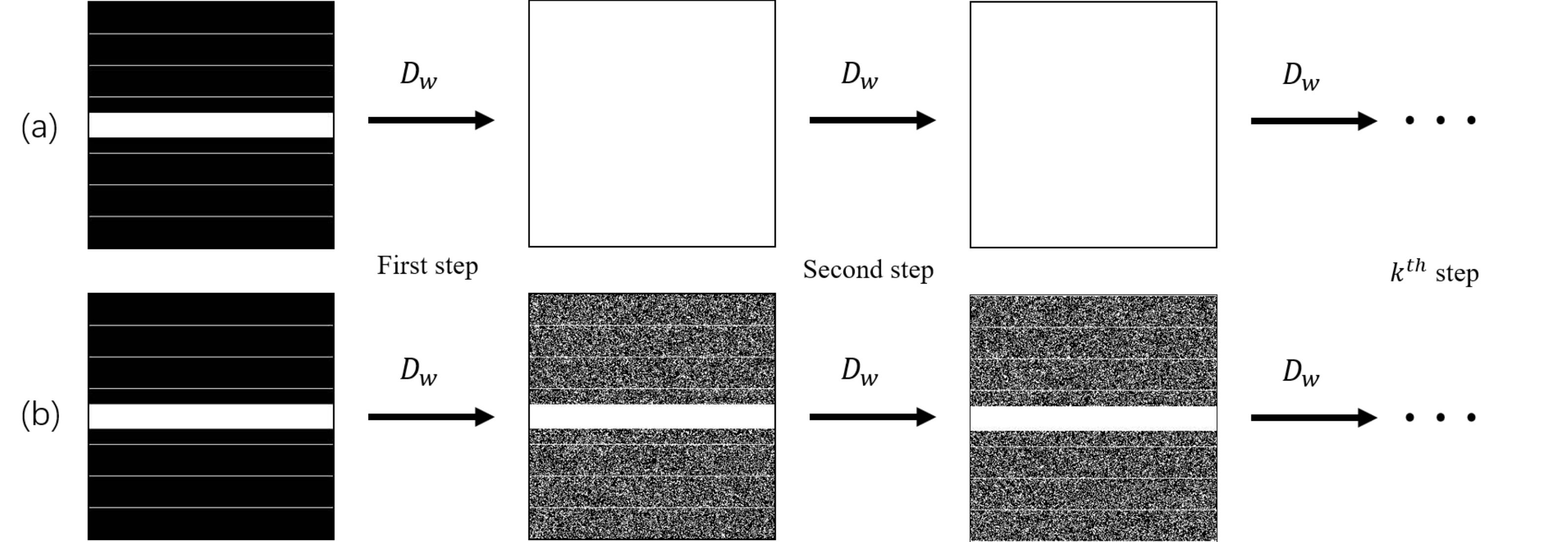}
\caption{Utilization of the learned K-space in iterative steps. The shown mask will be multiplied with the corresponding K-space measurements in practice. (a) Existing model-driven based reconstruction methods directly use the learned k-space with high noise ratio for the next iteration (Mask is represented as a blank matrix); and (b) the proposed K2Calibrate uses only part of the reconstructed signal.}
\label{fig_kt}
\end{figure}
Magnetic resonance imaging (MRI) is an important tool in the clinic for the detection of various diseases. One bottleneck issue of MRI is its time-consuming data acquisition process. Much effort has been devoted to alleviating this problem, among which compressed sensing (CS) based MR reconstruction has been acknowledged as a very effective approach. It exploits incoherent undersampling and image sparsity  \cite{lustig2007sparse,haldar2010compressed,trzasko2008highly} to accelerate MR imaging. Unfortunately, despite the promising performances reached with CS-MRI, it has several limitations such as the time-consuming iterative reconstruction process and the difficulty for the determination of its sparsity weight hyper-parameters\cite{shahdloo2018projection,ramani2012regularization}. 

To address these issues, deep learning (DL) has been introduced to accelerate MR imaging. One major group of methods are based on the data-driven approach. These methods directly learn a mapping between undersampled k-space data/aliased-image and fully sampled k-space data/artifact-free image in an end-to-end manner \cite{lee2018deep,wang2016accelerating,zhu2018image,mardani2018deep,han2019k}. Data-driven methods are straightforward and easy-to-implement, but they fail to exploit the physics of MRI. Model-driven methods, on the other hand, incorporate MRI physics into DL-based MRI reconstruction\cite{hammernik2018learning,modl,ista,admm,qin2018convolutional}. These methods unroll the iterative reconstruction steps of solving physical models into the different layers of a to-be-trained neural network \cite{liang2020deep,hammernik2018learning,modl,schlemper2017deep}. In this way, model-driven methods can integrate the strength of both DL and physical models. They have gradually become the prevalent solution for DL-based MRI reconstruction tasks.

Both data-driven and model-driven DL MRI reconstruction methods have made great progress in accelerating MR imaging. However, their performances rely on fully sampled and high quality reference datasets, the acquisition of which is not always feasible. One major reason is that fully sampled data acquisition is very time-consuming. For moving objects (such as the heart), high quality fully sampled dataset is hard to obtain without  image artifacts. For the same reason, fully sampled data are difficult to obtain when serious signal attenuation happens in some applications. For example, when conducting diffusion MRI scanning with echo planar (EPI), the T2 signal decays quickly. Thus, it is hard to adopt fully sampled data acquisition protocols. Another reason is acquiring a large number of fully sampled data to train a DL model is both resource- and labor-intensive.

Recently, Yaman et al.\cite{ssdu,yaman2021zero,yaman2021ground} proposed a series of MRI reconstruction methods based on model-driven self-supervised learning (self-supervised learning via data undersampling, SSDU), which does not require fully sampled reference data for model training. The core idea of SSDU is to split the obtained undersampled k-space data into two disjoint sets. One set is used as the input to the unrolled network, and the other set is used to supervise the training of the network. Later, Hu et al.\cite{hu} proposed an improved method based on SSDU to constrain the reconstruction results on non-sampled data points via a parallel network framework. Cole et al.\cite{gan} employed Generative Adversarial Networks (GAN) to guide the optimization of the reconstruction network via discriminating simulated and observed measurements. The above methods have achieved promising results using self-supervised MRI reconstruction methods. However, since the reconstruction problem is ill-posed due to sub-Nyquist sampling, imposing precise constraints on the non-sampled K-space data points is difficult. In addition, error propagation might happen alongside the iterative reconstruction process, especially at the initial iterations.

Here, we propose K2Calibrate, a convolutional neural network (CNN) based K-space adaptation method for self-supervised unrolled networks. By sampling k-space measurements from a Bernoulli distribution in each iteration step, the model-driven reconstruction process is optimized, so as to suppress the noise ratio caused by unsatisfied constraints of self-supervised models. The proposed method has high flexibility and can be formulated in a "plug-and-play" way , which can be easily applied to existing self-supervised model-driven methods without any complex operations.
\section{Theory}
Let $x$ denote the recovered image and $y$ represent the acquired k-space measurements. MRI image recovery using model-driven methods can be formed as the following optimization scheme:
\begin{equation}
\label{eq1}
\mathop{\arg\min}\limits_{x} \parallel Ax-y \parallel^2 +\lambda R(x)
\end{equation}
where $A=SFC$ is the encoding matrix including coil sensitivities $C$, 2-D discrete Fourier transform $F$, and 
sampling matrix $S$. The first term represents data consistency (DC) with acquired measurements. $R(\cdot)$ represents a regularization term. $\lambda$ denotes the weighting parameter. 
\subsection{Model based deep learning network }
Model based deep learning networks \cite{modl,admm,ista} have received extensive attention due to the enhanced physical interpretability. One typical example is MoDL \cite{modl}. Let $D_w$ represent a learned CNN estimator of noise depending on the parameters $w$. $x-D_w(x)$ can be treated as a denoiser to remove alias artifacts and noise. Then, Eq.\ref{eq1} can be rewritten as:
\begin{equation}
\mathop{\arg\min}\limits_{x} \parallel Ax-y \parallel^2 +\lambda \parallel(x-D_w(x))\parallel^2
\end{equation}

Since the non-linear mapping $D_w(x_k + \triangle x)$ can be approximated using Taylor series \cite{modl} in the unrolled networks, the reconstructed optimization problem can be approximated as:
\begin{equation}
\begin{array}{c}
x^{k+1}=\mathop{\arg\min}\limits_{x} \parallel Ax-y \parallel^2 +\lambda \parallel(x-z^k(x))\parallel^2 \\
z^k = D_w(x^k)
\label{eq3}
\end{array}
\end{equation}
Here, index $n \in [0, ..., K]$ denotes the iteration number. Once $K$ is fixed, the optimization process can be intuitively viewed as an unrolled linear CNN. In MoDL\cite{modl}, the authors used the same denoising operator $D_w$ and trainable regularization parameters at each iteration to reduce the network parameters. Since MoDL is a highly effective method, we adopt it as the baseline for our proposed method.

\subsection{Self-supervised MRI reconstruction}
\label{self-supervised}
\begin{figure}
\centering
\includegraphics[width=0.7\textwidth]{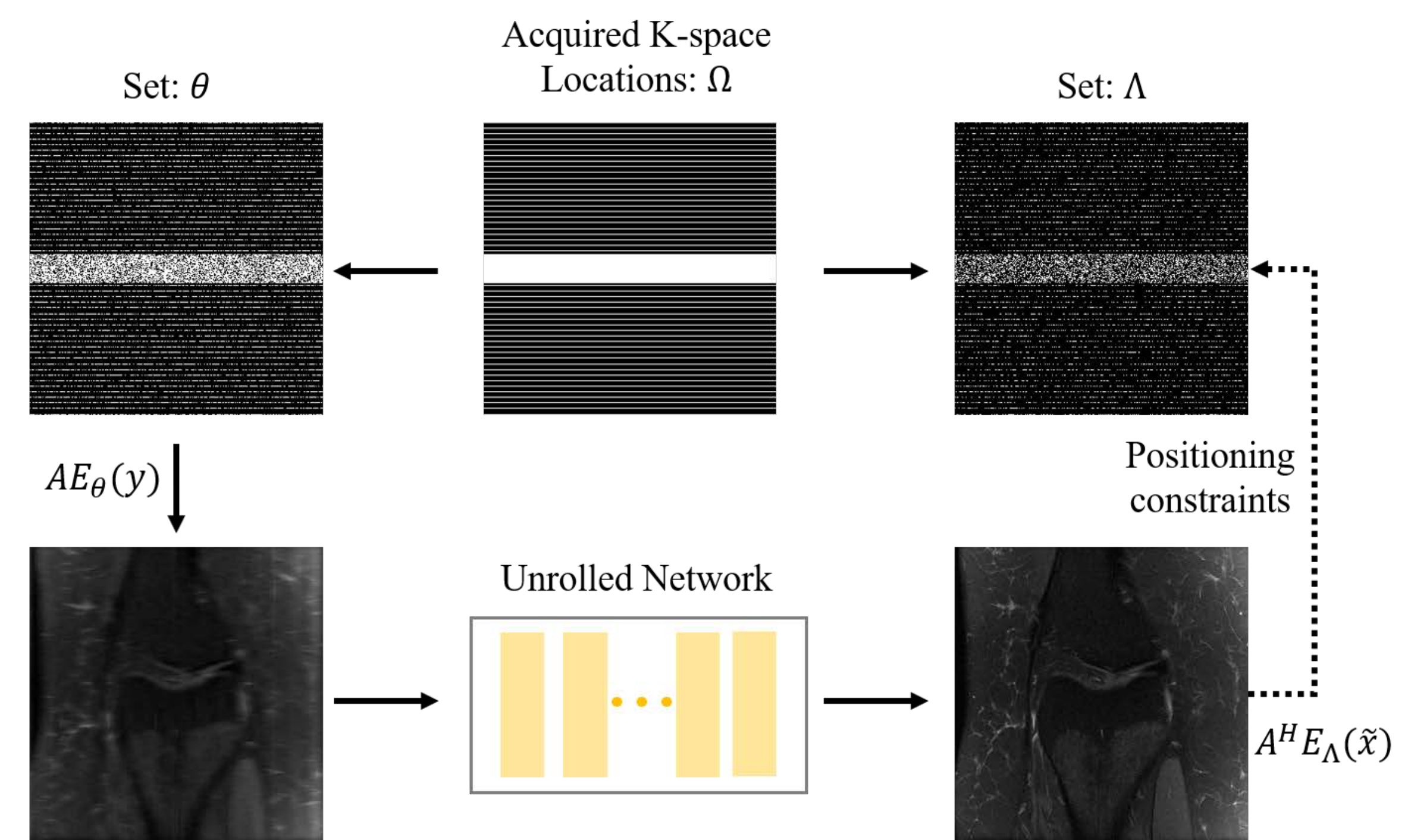}
\caption{The self-supervised MRI reconstruction workflow. The acquired k-space is split into sets $\Theta$ and $\Lambda$, while one is used for input and the other is used for supervision.}
\label{fig_process}
\end{figure}
Due to the difficulty of acquiring fully sampled reference data, Yaman et al.\cite{ssdu} proposed a self-supervised approach called SSDU achieving promising reconstruction performance. 

Let $\Omega$ denote the undersampled K-space measurements of fully sampled measurements $\Gamma$. For each data, $\Omega$ is divided into two random subsets, $\Theta$ and $\Lambda$, and the relationship between them is $\Omega = \Theta \cup \Lambda$.

To train the self-supervised model-driven network, the set of $\Theta$ is used as the input to the network as well as the DC units, while the set $\Lambda$ is used to define the similarity self-supervised loss function:
\begin{equation}
\label{eq4}
\mathop{\arg\min}\limits_{\omega} \emph{\textbf{L}}(\textbf{E}_\Lambda(y_\Omega),\textbf{E}_\Lambda(\emph{\textbf{f}}(\textbf{E}_\Theta(y);\omega)))
\end{equation}
where $\emph{\textbf{L}}(\cdot)$ is an arbitrary similarity function such as 1-norm or 2-norm and $\emph{\textbf{f}}(;\omega)$ denotes the output of the unrolled network with parameters $\omega$. $\textbf{E}_{\Lambda}(\cdot)$ represents a K-space transform operator that sampling the measurements  in $\Lambda$ region, and so as to $\textbf{E}_{\Theta}(\cdot)$. For clarification, let $y$ mentioned in equations (1,2,3) be $y_\Omega$ in this section.
In this way, a self-supervised strategy is implemented. This method updates the iterative network in an efficient end-to-end training manner. 

However, since the data are simulated by sub-Nyquist sampling, it inherently cannot constrain the reconstruction on the non-sampled data points $\Gamma \backslash \Omega$ precisely. This means once the reconstruction errors occur, they will accumulate alongside the unrolled iterative reconstruction process, as shown in Fig.\ref{fig_kt}(a). Unfortunately, according to Eq.\ref{eq4}, such reconstruction errors are easy to occur since there are inherently no precise supervision on $y_{\Gamma \backslash \Omega}$. 

\subsection{Derivation of K-space calibration}

\begin{figure}
\centering
\includegraphics[width=0.8\textwidth]{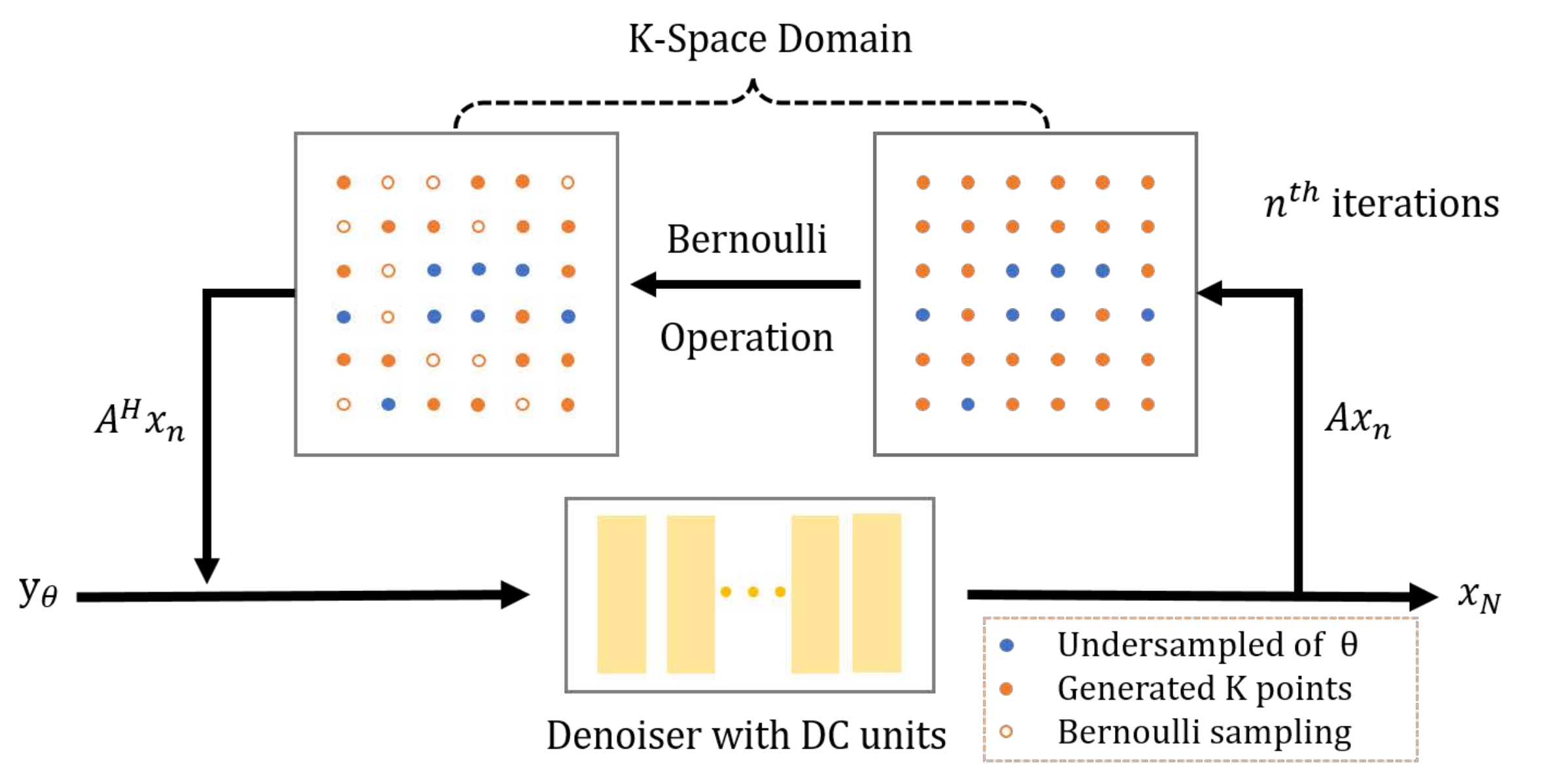}
\caption{The proposed K2Calibrate method. In each iteration, the sampled K-space measurements of $\theta$ are kept while the generated measurements from the network are sampled from Bernoulli distribution.}
\label{fig1}
\end{figure}

To analyze the possible impact of reconstruction errors, we now consider the output of a certain step of the unrolled networks $\tilde{x}^k=n^k+x^{*k}$. For simplicity, we omit the superscript $k$ and draw the joint distribution as follows:

\begin{equation}
p(x^*,n)=p(x^*)p(n\left|x^*\right)
\end{equation}

Here, $n$ denotes the noise randomly generated from the unrolled network while $x^*$ is the goal of the network in each specific iteration step. Due to the imaging physics of MRI, we assume that $x^*$ is statistically dependent that $p(x^*_i\left|x^*_j\right)\neq p(x^*_i)$. 

Then, we turn to consider the characteristics of the noise $n$. Since the task is based on self-supervised learning as mentioned in \ref{self-supervised}, $n$ becomes more diverse and should be considered more carefully. \cite{krull2019noise2void, lehtinen2018noise2noise} assume that $n$ of natural image is conditional independent such that  $p(n\left|x^*\right) = \prod p(n_i\left|x_i^*\right)$. By further assuming $\mathbb{E}[n_i]=0$, $n$ can be removed by using strategy of $\mathbb{E}[\tilde{x}_i]=x^*$. However, we believe $n$ is still dependent on $x^*$ due to the iterative process in MRI reconstruction that is more difficult to remove.

Here, we propose a K-space calibration strategy (Fig.\ref{fig1}), called 'K2Calibrate', to try to mitigate the effect of noise in the unrolled networks. Different from designing a complexity prior of $n$ manually, we speculate that the reconstruction problem can be transferred to balance $n$ and $x^*$ during the iteration process. Let's divide $n$ into two parts, $n_s$ and $n_g$, denoting noise at location of sampled/non-sampled K-space measurements, respectively. Since DC layers are included in the unrolled networks, $\mathbb{E}[n_s]$ is closed to 0 and is much smaller than $\mathbb{E}[n_g]$. Therefore, if we choose a tiny increment $n_\delta \in n_g$, we have:
\begin{equation}
\mathbb{E}[\tilde{x}_{s+\delta}]=\mathbb{E}[x^*_{s+\delta}+n_s+n_\delta]\approx x^*_{s+\delta}
\end{equation}

which significantly reduces the scale of $n$ in $\tilde{x}$ comparing to $\mathbb{E}[\tilde{x}_{s+g}]=x^*_{s+\delta}+n_\delta$. Based on this observation, we can prevent the accumulation of noise in each iteration by reducing the size of $n$ generated from the previous iteration. We modified Eq.\ref{eq3} to achieve this and make sure that noise can be removed iteratively rather than all at once in each iteration:
\begin{equation}
\begin{array}{c}
x^{k+1}=\mathop{\arg\min}\limits_{x} \parallel Ax-y \parallel^2 +\lambda \parallel(x-z^k(x))\parallel^2 \\
z^k = \textbf{E}^k_{s+\delta}(D_w(x^k))
\end{array}
\end{equation}

In experiments, we use Bernoulli-sampling to simulate the incremental $\delta \in g$. By adapting this calibration strategy in K-space measurements, $D_w$ can be optimized to learn more robust representations since the calibration operation reduces the propagation of $n_g$ in the unrolled networks, as shown in Fig.\ref{fig_kt}(b), especially in the initial iteration steps.

\section{Experiments and Results}

\subsection{Dataset}
\label{data}
We use an open-source MR multi-coil knee dataset, FastMRI, to verify the effectiveness of the proposed method. FastMRI is a large-scale data collection of clinical MR measurements released by Facebook AI Research\cite{zbontar2018fastmri}. 1,594 scans were collected utilizing the following scanning parameters: Echo train length 4, 15-channels coils, matrix size of 320$\times$320, slice thickness 3mm, in-plane resolution 0.5mm×0.5mm, and no gap between slices. The dataset was split into training, validation, and testing sets. During training, we train the models with 3$\%$ (Set-A) and 1$\%$ (Set-B) data of the training set to evaluate the performance of different methods when different numbers of training data are utilized.

\subsection{Implementation details}
Multiple comparison methods are implemented. Among them, we use Sigpy\cite{ong2019sigpy} package to implement L1-Wavelet, SENSE, and TotalVariation (TV). DL-based self-supervised methods include SSDU\cite{ssdu} and Hu et al.\cite{hu} and supervised MoDL\cite{modl} are also experimented with. For fair comparison, all the CNN-based methods adopt the same network architecture proposed in \cite{modl}. $D_w$ block has 9 convolution layers (kernel no of 64). Each convolution layer is followed by a non-linear activation LeakyReLu except for the final layer. The output of each $D_w$ block is fed to a data consistency (DC) layer as described in \cite{modl}. By sharing the learning weight of $D_w$, we unroll the network into 5 and 10 iterations according to the experiments settings, and adopt the proposed K2Calibrate strategy at different steps in the iterations. It is worth noting K2Calibrate is only enabled during model training and it is removed during model testing. Adam optimizer is used with a learning rate of $5\times10^{-4}$ and a batch size of 2 for model training. All experiments are performed using Pytorch in Python.

\subsection{Results}
This section presents various quantitative and qualitative evaluations to verify the effectiveness of the proposed method. 
\subsubsection{Quantitative analysis} Two evaluation metrics, peak signal to noise ratio (PSNR) and structural similarity index measure (SSIM), are calculated to quantitatively evaluate the image reconstruction results. Table \ref{tab1} lists the results of conventional methods and DL-based methods. Both methods do not require ground truth fully sampled reference data for model training. Compared with the conventional methods, DL-based methods show better performance due to the complex denoiser provided by its large number of parameters. When the proposed K2Calibrate in utilized, the metrics of the two self-supervised DL-based methods improve obviously. SSIM is increased from 0.7674 to 0.7803 for SSDU(8x) and from 0.7680 to 0.7754 for Hu(8x), respectively. Similar performance improvement is observed for other settings, which validates K2Calibrate that can improve the performance of different reconstruction methods.

\begin{table}[htbp!]
\caption{Quantitative results on Set-A with acceleration rates of 4 and 8 of different self-supervised MRI reconstruction methods.} 
\label{tab1}
\centering
\begin{tabular}{|c|cl|cl|}
\hline
\multirow{2}{*}{Method} & \multicolumn{2}{c|}{4x}                                            & \multicolumn{2}{c|}{8x}                                            \\ \cline{2-5} 
                        & \multicolumn{1}{c|}{PSNR}            & \multicolumn{1}{c|}{SSIM}   & \multicolumn{1}{c|}{PSNR}            & \multicolumn{1}{c|}{SSIM}   \\ \hline
L1Wavelet               & \multicolumn{1}{c|}{32.172}          & \multicolumn{1}{c|}{0.7674} & \multicolumn{1}{c|}{28.704}          & \multicolumn{1}{c|}{0.6749} \\
SENSE                   & \multicolumn{1}{c|}{35.323}          & \multicolumn{1}{c|}{0.9121} & \multicolumn{1}{c|}{29.461}          & \multicolumn{1}{c|}{0.7636} \\
TV                      & \multicolumn{1}{c|}{31.856}          & \multicolumn{1}{c|}{0.7725} & \multicolumn{1}{c|}{29.361}          & \multicolumn{1}{c|}{0.7175} \\ \hline
SSDU                    & \multicolumn{1}{l|}{37.795}          & 0.9332                      & \multicolumn{1}{l|}{29.584}          & 0.7674                      \\
SSDU-K2C                & \multicolumn{1}{l|}{\textbf{38.496}} & \textbf{0.9402}             & \multicolumn{1}{l|}{\textbf{30.303}} & \textbf{0.7803}             \\
Hu                      & \multicolumn{1}{l|}{37.670}          & 0.9341                      & \multicolumn{1}{l|}{29.604}          & 0.7680                      \\
Hu-K2C                  & \multicolumn{1}{l|}{\textbf{37.947}} & \textbf{0.9361}             & \multicolumn{1}{l|}{\textbf{29.841}} & \textbf{0.7754}             \\ \hline
\end{tabular}
\end{table}

We implemented our method on the smaller training set, Set-B, and results are listed in Table \ref{tab2}. Since the conventional methods are not affected by the number of training data, the performance of them is the same as that presented in Table \ref{tab1}. Compared with Table \ref{tab1}, the overall scores in Table \ref{tab2} are slightly decreased due to that fewer training data are utilized. However, on this smaller-scale dataset, K2Calibrate can still improve the model performance, indicating that the effectiveness of K2Calibrate is not dependent on the amount of training data.

\begin{table}[htbp!]
\caption{Quantitative results on Set-B with acceleration rates of 4 and 8. K2Calibrate can still improve the reconstruction performance of the two self-supervised methods.} 
\label{tab2}
\centering
\begin{tabular}{|c|cc|cc|}
\hline
\multirow{2}{*}{Method} & \multicolumn{2}{c|}{4x}                                & \multicolumn{2}{c|}{8x}                                \\ \cline{2-5} 
                        & \multicolumn{1}{c|}{PSNR}            & SSIM            & \multicolumn{1}{c|}{PSNR}            & SSIM            \\ \hline
SSDU                    & \multicolumn{1}{c|}{36.936}          & 0.9242          & \multicolumn{1}{c|}{29.447}          & 0.7634          \\
SSDU-K2C                & \multicolumn{1}{c|}{\textbf{37.693}} & \textbf{0.9313} & \multicolumn{1}{c|}{\textbf{29.783}} & \textbf{0.7735} \\
Hu                      & \multicolumn{1}{c|}{36.926}          & 0.9256          & \multicolumn{1}{c|}{29.600}          & 0.7681          \\
Hu-K2C                  & \multicolumn{1}{c|}{\textbf{37.356}} & \textbf{0.9301} & \multicolumn{1}{c|}{\textbf{29.642}} & \textbf{0.7692} \\ \hline
\end{tabular}
\end{table}

\begin{figure}
\centering
\includegraphics[width=1.0\textwidth]{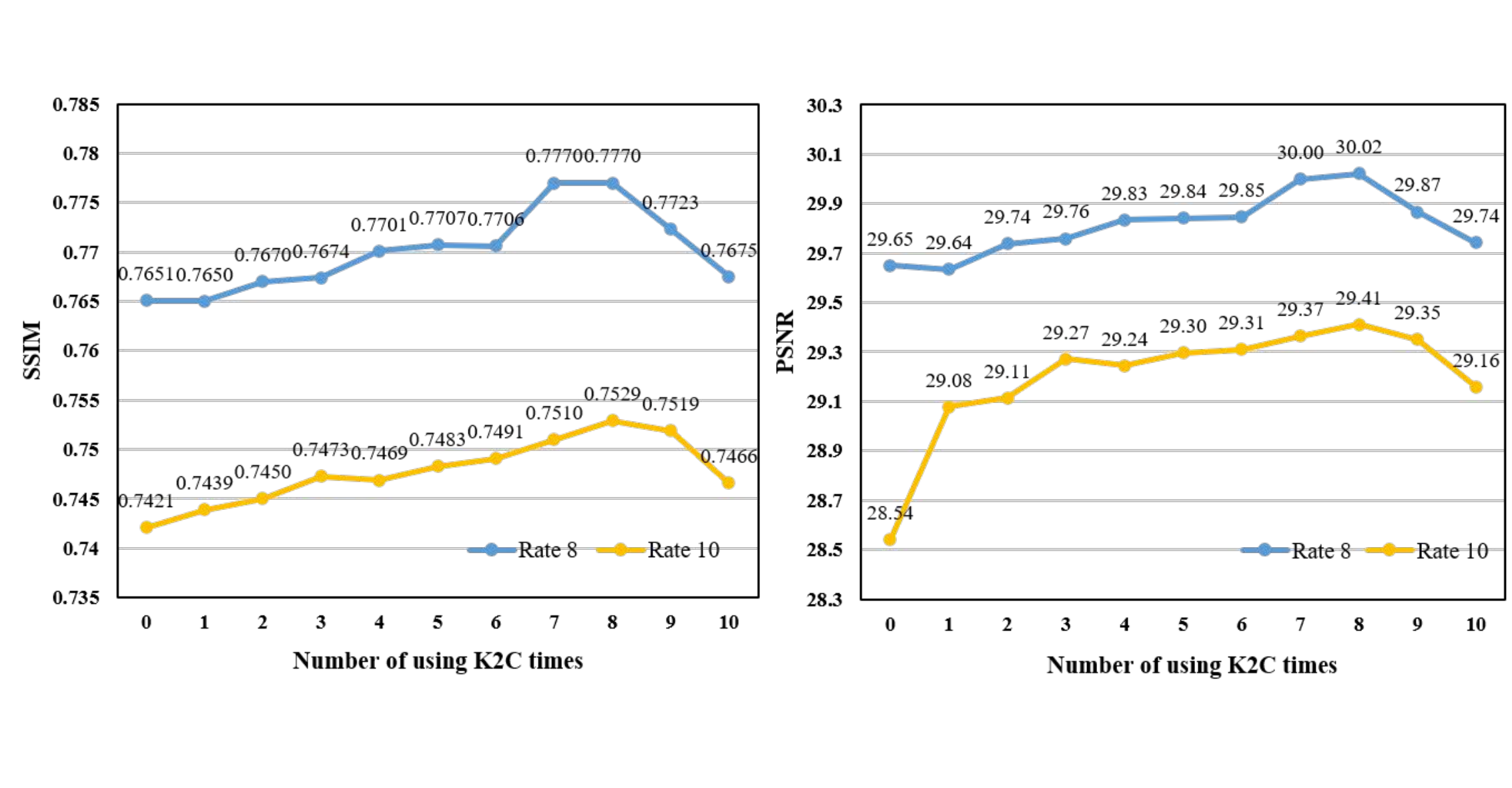}
\caption{Reconstruction results when different numbers of K2Calibrate are enabled. Acceleration rates of 8 and 10 are utilized. The number of K2calibrate indicate how many iterations utilized K2Calibrate. For example, if the number of K2Calibrate is 5, it means K2Calibrate is enabled in only the first 5 training iterations.} 
\label{fig_plot}
\end{figure}

K2Calibrate is a plug-and-play module. To have a clear understanding of the impact of K2Calibrate on the reconstruction performance, we have conducted experiments by enabling K2Calibrate in different network training iterations. SSDU was adopted as the base architecture in this set of experiments. Relatively higher acceleration rates (8x and 10x) were employed to better visualize the improvements made by K2Calibrate. Results are plotted in Fig. 4. In the first 8 iterations, the reconstruction performance is enhanced when more K2Calibrate modules are enabled. Compared to the method without K2Calibrate, utilizing K2Calibrate in the first iteration increases the PSNR value significantly from 28.54 db to 29.08 db under the acceleration rate of 10, which validates the effectiveness of K2Calibrate on noise suppression. When more than 8 K2Calibrate modules are enabled (the number of K2calibrate is 9 or 10 in Fig.\ref{fig_plot}), the performance becomes worse. This observation is in accordance with our expectations. We have explained that the noise ratio is relatively higher in the initial iterations, and thus utilizing K2Calibrate in these iterations is essential for noise removal. After these iterations, utilizing more K2Calibrate modules becomes ineffective.
\begin{figure}
\centering
\includegraphics[width=0.7\textwidth]{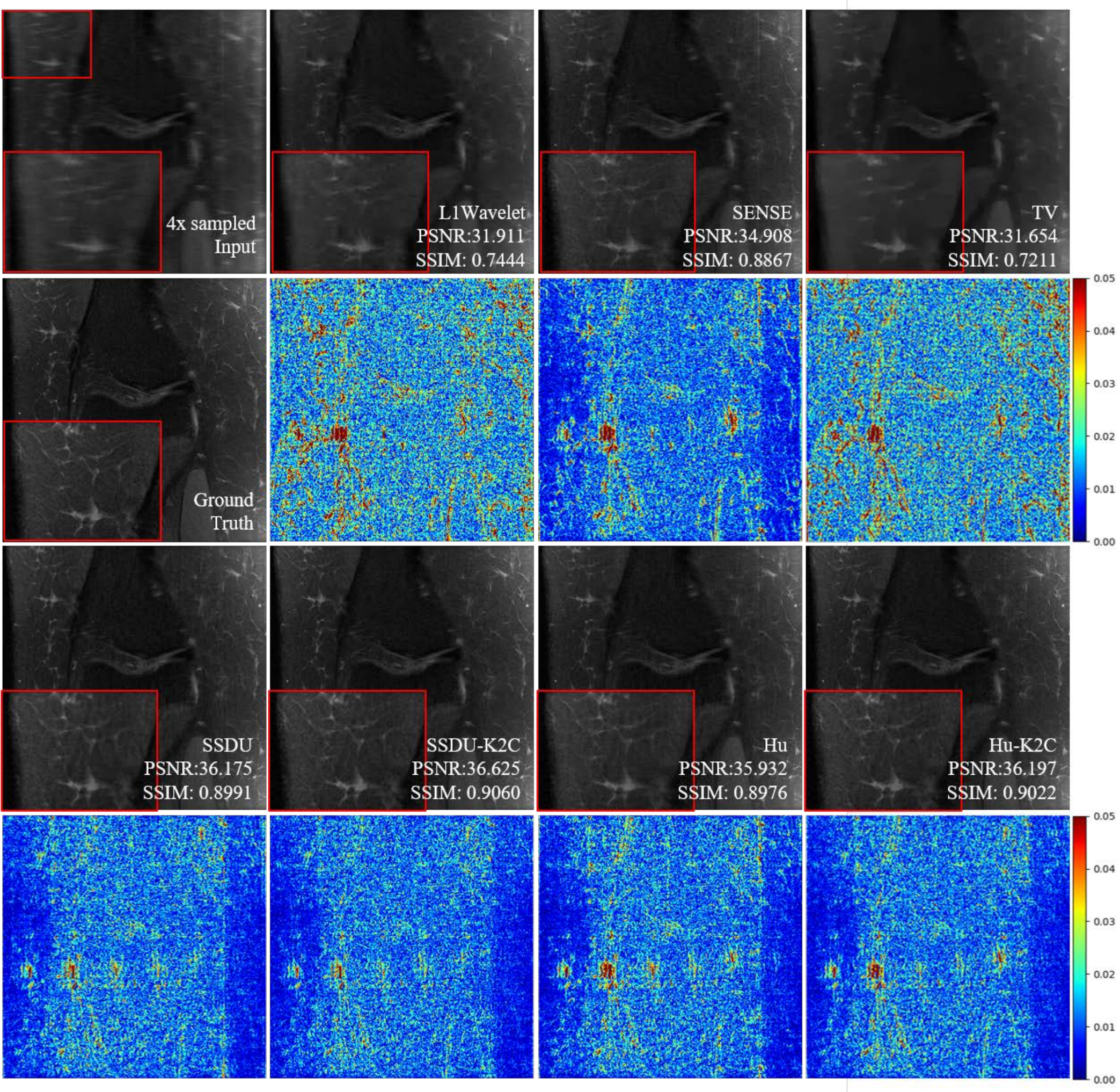}
\caption{Qualitative results with the acceleration rate of 4. Blue graph represents the differences between the fully-sampled image (Ground Truth) and the reconstructed image.}
\label{fig_res}
\end{figure}
The above experiments show that self-supervised MRI reconstruction methods can outperform the conventional methods and our method can effectively improve the performance of the self-supervised model-driven methods regardless of the acceleration rates utilized. By experimenting with Set A and Set B, we prove that the performance improvements of the proposed method is not dependent on the size of the training set. Extensive experiments have been conducted by utilizing different numbers of K2Calibrate modules during model optimization with higher acceleration rates of 8x and 10x. Results confirm that K2Calibrate is effective for noise removal in initial iterations. All these experimental results validate the performance enhancement of the proposed K2Calibrate for self-supervised DL-based MRI reconstruction.

\subsubsection{Qualitative Analysis}

Fig.\ref{fig_res} depicts the reconstructed images under the acceleration rate of 4 of conventional and DL-based methods. Based on the residual maps, the conventional methods suffer from extensive loss of structure details while both DL-based methods can recover these details more accurately, showing that DL-based methods are more effective for the recovery of high-frequency measurements. Building on two state-of-the-art self-supervised deep learning methods, K2Calibrate can further improve the reconstruction performance with fewer errors observed in the residual maps. These results demonstrate that the proposed method can successfully remove the residual artifacts, while achieving higher qualitative and quantitative performances compared to the baseline methods.

\section{Conclusion}
In this study, we provide a fresh perspective to review the optimization process of iterative DL-based reconstruction. We propose K2Calibrate, which is a plug-and-play module that help the unrolled network gradually reconstruct K-space measurements. Compared to the existing self-supervised learning methods that reconstruct all the K-space measurements at once with high uncertainties and propagate them to the following network training iterations, K2Calibrate can effectively reduce the error accumulation. Experiments on the FastMRI dataset confirm that the proposed method can achieve better results than state-of-the-art self-supervised methods even the same network architecture is used. Overall, K2Calibrate can improve the performance of self-supervised reconstruction methods effectively and can be easily plugged into different model-driven DL-based methods without any additional operations.

\clearpage
%
%
\bibliographystyle{unsrt}
\bibliography{mybibliography}
\end{document}